\documentclass[12pt]{aastex631}

\pdfoutput=1

\shorttitle{Optical Flares of Wolf\,359}
\shortauthors{Lin et al. }

\begin{document}

\title{
       Simultaneous Detection of Optical Flares of the Magnetically Active M Dwarf Wolf\,359  
      }

\correspondingauthor{Wen Ping Chen}
\email{wchen@astro.ncu.edu.tw}

\author{Han-Tang Lin}
\affiliation{Graduate Institute of Astronomy, 
National Central University, 300 Zhongda Road, Zhongli, Taoyuan 32001, Taiwan}

\author[0000-0003-0262-272X]{Wen-Ping Chen}
\affiliation{Graduate Institute of Astronomy, 
  National Central University, 300 Zhongda Road, Zhongli, Taoyuan 32001, Taiwan}
\affiliation{Department of Physics, 
  National Central University, 300 Zhongda Road, Zhongli, Taoyuan 32001, Taiwan}

\author[0000-0002-7420-6744]{Jinzhong Liu}
\affiliation{Xinjiang Astronomical Observatory, Chinese Academy of Sciences, People's Republic of China}

\author[0000-0002-5750-8177]{Xuan Zhang}
\affiliation{Xinjiang Astronomical Observatory, Chinese Academy of Sciences, People's Republic of China}

\author[0000-0001-7134-2874]{Yu Zhang}
\affiliation{Xinjiang Astronomical Observatory, Chinese Academy of Sciences, People's Republic of China}

\author[0000-0003-4708-5964]{Andrew Wang}
\affiliation{Institute of Astronomy and Astrophysics, Academia Sinica, No.~1, Sec.~4, Roosevelt Rd, Taipei 10617, Taiwan}

\author[0000-0001-6491-1901]{Shiang-Yu Wang}
\affiliation{Institute of Astronomy and Astrophysics, Academia Sinica, No.~1, Sec.~4, Roosevelt Rd, Taipei 10617, Taiwan}

\author[0000-0003-4077-0985]{Matthew J. Lehner}
\affiliation{Institute of Astronomy and Astrophysics, Academia Sinica, No.~1, Sec.~4, Roosevelt Rd, Taipei 10617, Taiwan}
\affiliation{Department of Physics and Astronomy, University of Pennsylvania, 209 South 33rd Street, Philadelphia, PA 19125, USA}
\affiliation{Center for Astrophysics, Harvard \& Smithsonian, 60 Garden Street, Cambridge, MA 02138, USA}

\author{C. Y. Wen}
\affiliation{Institute of Astronomy and Astrophysics, Academia Sinica, No.~1, Sec.~4, Roosevelt Rd, Taipei 10617, Taiwan}

\author{J. K. Guo}
\affiliation{Graduate Institute of Astronomy, 
  National Central University, 300 Zhongda Road, Zhongli, Taoyuan 32001, Taiwan}

\author{Y. H. Chang}
\affiliation{Graduate Institute of Astronomy, 
  National Central University, 300 Zhongda Road, Zhongli, Taoyuan 32001, Taiwan}

\author{M. H. Chang}
\affiliation{Graduate Institute of Astronomy, 
  National Central University, 300 Zhongda Road, Zhongli, Taoyuan 32001, Taiwan}

\author{Anli Tsai}
\affiliation{Graduate Institute of Astronomy, 
  National Central University, 300 Zhongda Road, Zhongli, Taoyuan 32001, Taiwan}

\author[0000-0001-5989-7594]{Chia-Lung Lin}
\affiliation{Graduate Institute of Astronomy, 
  National Central University, 300 Zhongda Road, Zhongli, Taoyuan 32001, Taiwan}
  
\author{C. Y. Hsu}
\affiliation{Graduate Institute of Astronomy, 
  National Central University, 300 Zhongda Road, Zhongli, Taoyuan 32001, Taiwan}

\author{Wing Ip}
\affiliation{Graduate Institute of Astronomy, 
  National Central University, 300 Zhongda Road, Zhongli, Taoyuan 32001, Taiwan}
\affiliation{Graduate Institute of Space Science, 
  National Central University, 300 Zhongda Road, Zhongli, Taoyuan 32001, Taiwan}

\begin{abstract}
We present detections of stellar flares of Wolf\,359, an M6.5 dwarf in the solar neighborhood (2.41~pc) known to be prone to flares due to surface magnetic activity. The observations were carried out from 2020 April 23 to 29 with a 1-m and a 0.5-m telescope separated by nearly 300~km in Xinjiang, China.  In 27~hr of photometric monitoring, a total of 13 optical flares were detected, each with a total energy of $\gtrsim 5 \times 10^{29}$~erg. The measured event rate of about once every two hours is consistent with those reported previously in radio, X-ray and optical wavelengths for this star. One such flare, detected by both telescopes on 26 April, was an energetic event with a released energy of nearly $10^{33}$~erg.  The two-telescope lightcurves of this major event sampled at different cadences and exposure timings enabled us to better estimate the intrinsic flare profile, which reached a peak of up to 1.6 times the stellar quiescent brightness, that otherwise would have been underestimated in the observed flare amplitudes of about $0.4$ and $0.8$, respectively, with single telescopes alone.  The compromise between fast sampling so as to resolve a flare profile versus a longer integration time for higher photometric signal-to-noise provides a useful guidance in the experimental design of future flare observations.
\end{abstract}


\section{Introduction} \label{sec:int}

Solar flares are commonly observed surface phenomena, attributable to acceleration of plasma during magnetic reconnection near sunspots and active regions, that lead to sudden brightening observed in radio, optical, to X-ray wavelengths.  While the detailed heating mechanism, i.e., how the magnetic energy is converted to gas kinetic energy is still unclear \citep{ben10}, it is believed that conductive and radiative processes are involved in the cooling phase.   The solar flares are classified by the peak flux in soft X-rays, with the most powerful being class X peaking at $>0.1$~erg~s~$^{-1}$~cm$^{-2}$.  A typical solar flare releases $10^{29}$--$10^{32}$~erg.  Rare, major flares, which release a total energy more than $\sim10^{33}$~erg, are linked to coronal mass ejection events which influence space weather and pose potential hazards to terrestrial environments.  

Other stars, notably late-type dwarfs, being largely convective are even more predisposed to flare activity and encompass a larger range of energy output, particularly the young ones with fast rotation rates \citep{fei20}.  Based on the K2 lightcurves of G- to M-type dwarfs, \citet{lin19} conclude that later type stars have higher flare occurrence frequencies but generally with less energetic output.  M-dwarf flares with much shorter durations have been detected in millimeter wavelengths \citep{mac20}. Later spectral types (brown dwarfs) have also shown surface activity \citep[e.g.,][]{rut00}.  

Our target, Wolf\,359 (GJ~406; CN\,Leo), at a heliocentric distance of 2.41~pc \citep{gai21}, is a known eruptive-type red subdwarf \citep{kes19}. Previously, six flare events were detected during 12.8~hours of monitoring in radio frequencies, equivalent to 47 events per 100 hours \citep{nel79}.  Extreme ultraviolet flare events have been reported to occur at least daily \citep{aud00}, and major X-ray flares have also been detected in this star \citep{lie10}. Using ground-based and Kepler/K2 observations combining long- and short-cadence lightcurves of Wolf\,359, \citet{lin21} derived a flare occurrence rate of once per day for events with a total flare energies $>10^{31}$~erg, and ten times per year for superflares with released energies $>10^{33}$~erg. Such an activity level is considered high even among known flaring M dwarfs.  

Magnetic reconnection may not be limited to surface extrusion as in the case of the Sun.  The field lines may be linked to some intricate spin-orbit magnetospheric interaction with close-in companion stars, as in RS Canum Venaticorum or BY Draconi type variables.  It is known that newly born stars may anchor their magnetic field to circumstellar disks \citep{fei99}, with which the entwined field lines are susceptible to reconnection and result in outbursts ($10^{36}$~erg) or in extended flaring loops \citep{hay96,shi11}.  Superflare events of red dwarfs are suspected to have similar interactions with orbiting giant exoplanets \citep[e.g.,][]{klo17}, though there is so far no definite supporting evidence. Wolf\,359 is a fast rotator \citep[][2.72~d]{gui18} and is known to host at least two exoplanets \citep{tuo19}. One of these, Wolf\,359c (radius 0.1272~R$_\mathrm{Jupiter}$) is hot and close in (0.018~au, orbital period 2.88~d)  suggestive of a possible spin-orbit tidal lock. In the flare star AD\,Leo, a periodicity of 2.23~d was inferred and attributed to stellar rotation \citep{hun12}.  However, \citet{lin21} found no flare timing in synchronization with the planetary orbital phase in Wolf\,359. 

Here we report on an optical monitoring campaign of Wolf\,359. The star was monitored for one week in 2020 April simultaneously with two telescopes. In addition to reaffirmation of the flare rate, with 13 events detected in 27 observing hours, this paper focuses on one major flare observed by both telescopes, affording the possibility to derive the underlined flare profile, whose amplitude would have been underestimated by any single lightcurve as a result of finite exposure time. While a flare is quantified by the total released energy (essentially scaled with the amplitude multiplied by the duration), usually an event is recognized mainly by a brightness spike.  Our study indicates how intrinsically moderate events could escape detection, and provides guidelines for proper sampling specific to certain profiles in the experimental design. 

\section{Observations and Data Analysis}

\subsection{CCD Imaging and Lightcurve Extraction} 

The observations reported here were carried out from 2020 April 23 to 29 simultaneously by the Nanshan One-meter Wide-field Telescope (NOWT) in Xinjiang, and one of the TAOS telescopes, except for the night of April 24 for which the TAOS site was weathered out.  The TAOS telescopes, each of f/2 50~cm, used to be installed at Lulin Observatory to catch chance stellar occultation events by transneptunian objects \citep[e.g., ][]{alc03,zha13}. Two of the original four TAOS telescopes were relocated in the spring of 2020 to Qitai Station in Xinjiang, some 300~km from Nanshan, also operated by Xinjiang Astronomical Observatory.

The NOWT was equipped with an E2V back-illuminated CCD203-82 camera, with 12-micron pixels, spanning 1\farcs13/pixel on the sky.  For the data reported here, the NOWT observed Wolf\,359 in the $R$ band for the first five nights, and in the $V$ band for the rest two nights.  The exposure time was 18~s, with a dead time of approximately 12~s between exposures, amounting to a cadence of $\sim30$~s.

The TAOS telescope was equipped with a Spectral Instrument 800 camera with  13.5-micron pixels and a plate scale of 2\farcs78/pixel.  A custom-made  filter was used which has a flat-response in 500--700~nm approximately comparable to an SDSS $r^\prime$ filter.  For the observing campaign of Wolf\,359, the exposure time was 45~s, with a dead time of $\sim0.5$~s.  

\begin{figure*}[htbp]
\centering
  \includegraphics[width=\textwidth,angle=0]{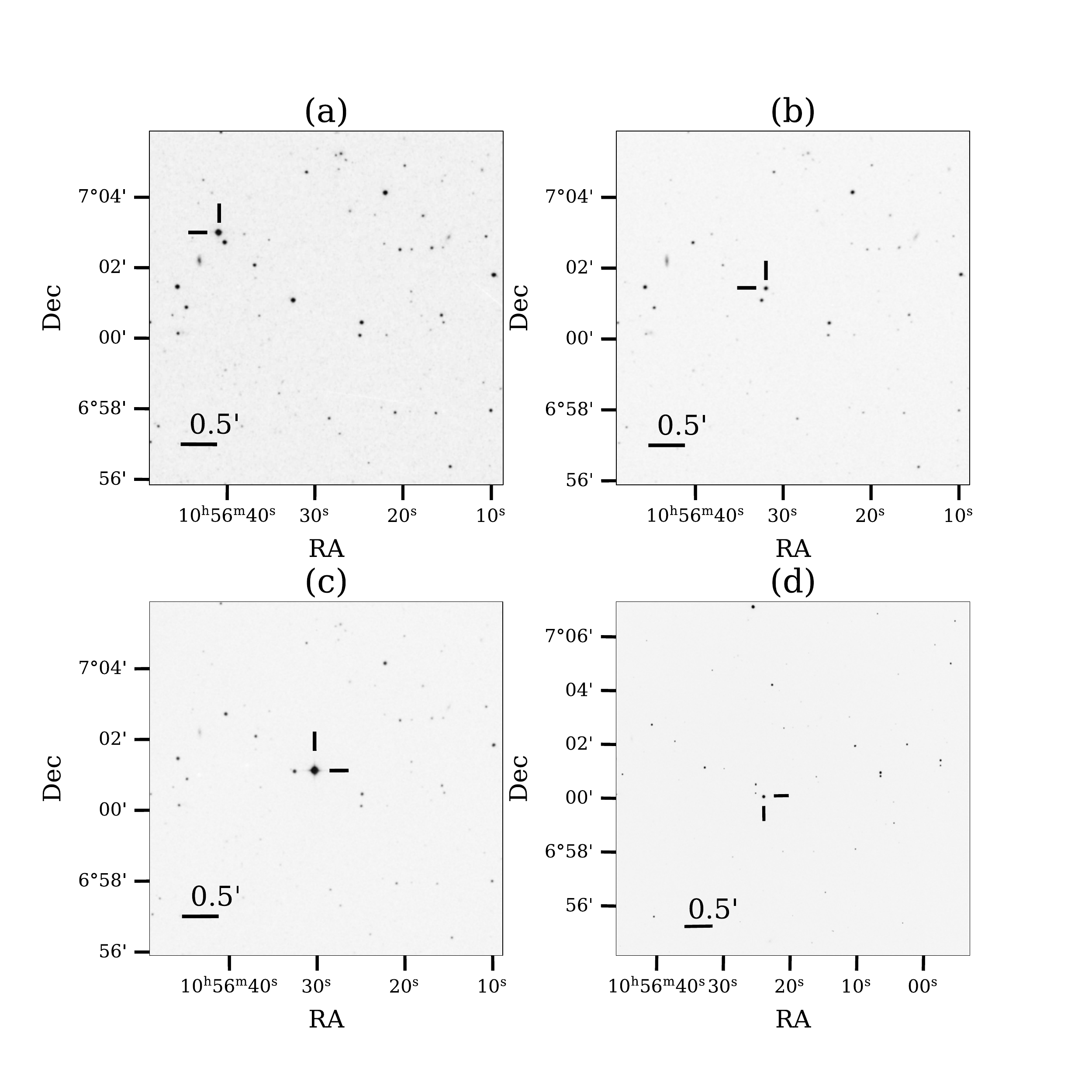}
  \caption{The motion of Wolf\,359.  The first three images are from Digital Sky Survey taken in (a) 1953, (b) 1988, and (c) 1995, whereas the last image (d) was taken in 2020 as a part of this work, all shown with J2000 coordinates. 
  }
  \label{fig:chart}
\end{figure*}

Wolf\,359 has a high proper motion due to its proximity, with $\mu_\alpha =-3866.338\pm0.081$~mas~yr$^{-1}$ and $\mu_\delta= -2699.215 \pm 0.069$~mas~yr$^{-1}$ \citep{gai21}.  Figure~\ref{fig:chart} displays its position in four epochs, three recorded by the Digital Sky Survey in years 1953, 1988 and 1995, whereas the last one was taken in 2020 reported in this work.

No attempt was made to synchronize the shutter openings of the two telescopes.  The different cadences, hence sampling functions, of the two telescopes observing the same flare event in turn provide the possibility to derive the underlined flare profile, which would not have been possible otherwise with a single data set.  

\begin{figure*}[htbp]
\centering
  \includegraphics[width=0.45\textwidth,angle=0]{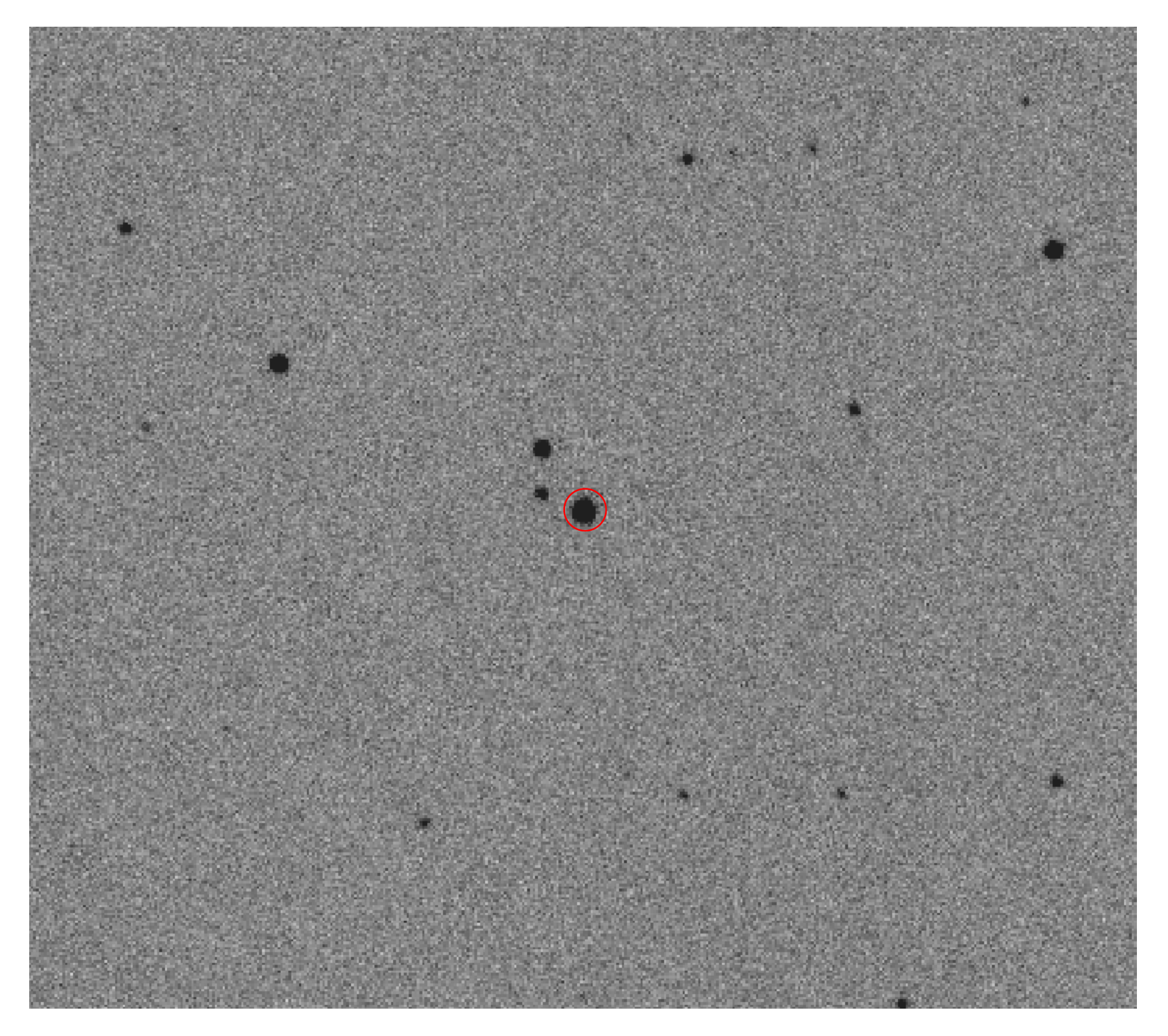}
  \includegraphics[width=0.47\textwidth,angle=0]{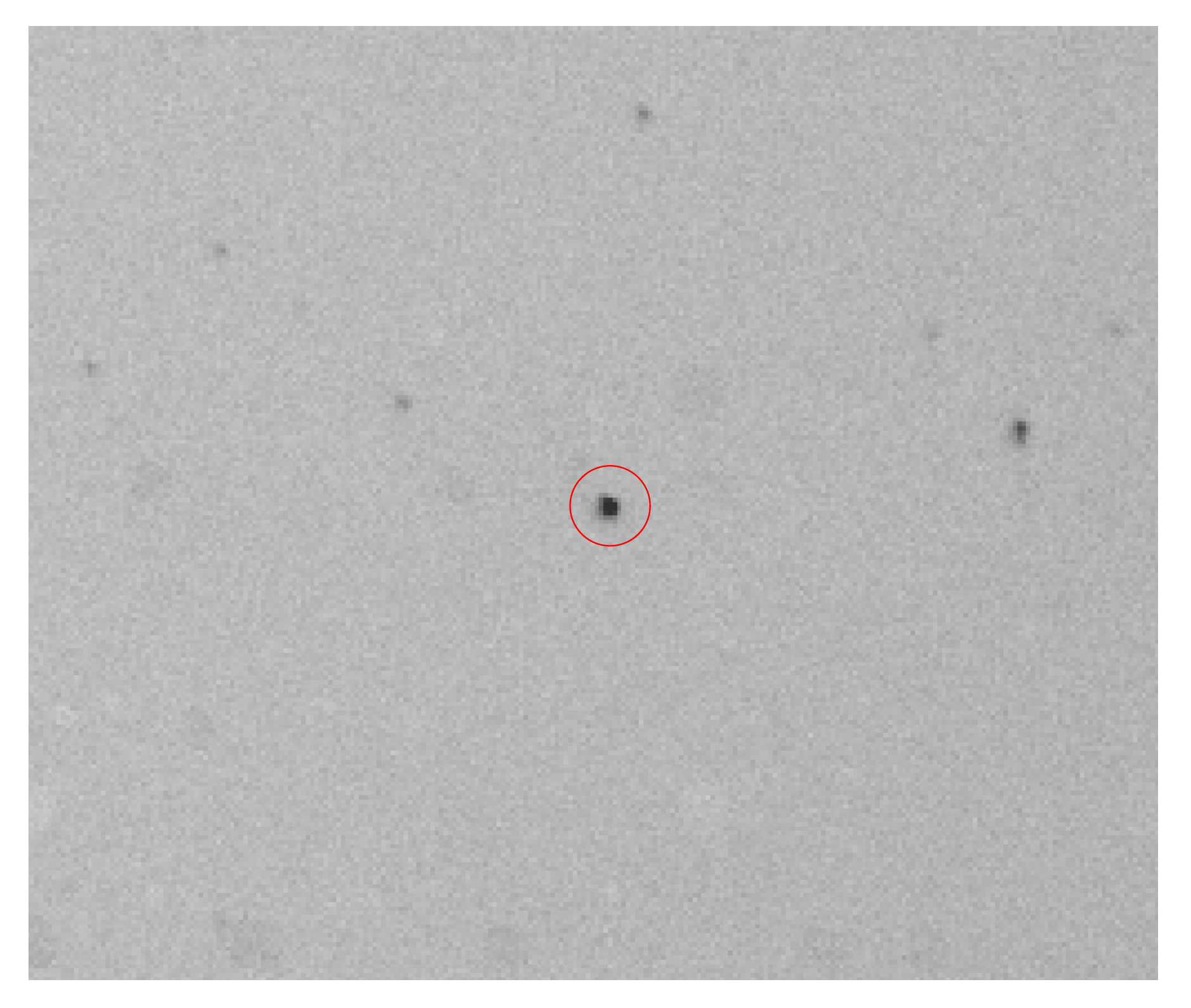}
  \caption{Example images with Wolf\,359 each marked with a red circle depicting the maximal photometric aperture size used in lightcurve extraction: 8 pixels for NOWT (left image), and 9 pixels for TAOS (right image). }
  \label{fig:twoTelCharts}
\end{figure*}

Images were processed by the standard procedure of bias, dark, and flat-field corrections. Aperture photometry was then performed using SExtractor \citep{ber96,ber10} with an adaptive aperture to measure the brightness of Wolf\,359 with an aperture size of 4--8 pixels for the NOWT images and of 8--9 pixels for the TAOS images.  Figure~\ref{fig:twoTelCharts} shows an illustrative image taken by NOWT and by TAOS reported here with the photometric aperture marked.

\begin{figure*}[htbp]
\centering
  \includegraphics[width=\textwidth,angle=0]{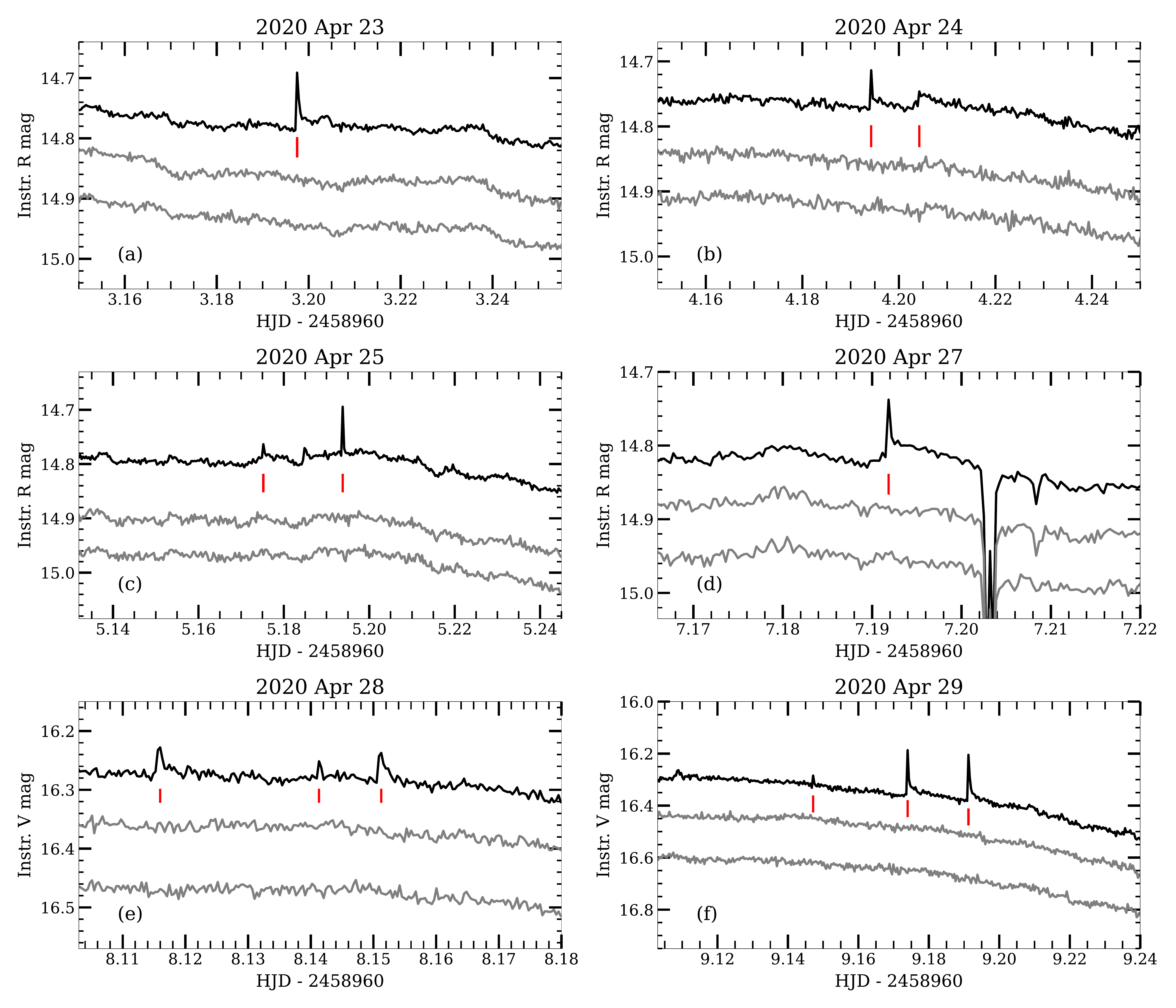}
  \caption{The NOWT lightcurves of Wolf\,359 (in black) and of the two reference stars (in gray).  Each flare event detected is marked by a vertical red line.  The observations from April 23 to 27 were taken in the $R$ band, whereas on April 28 and 29 they were in $V$.  The sudden flux drop on April 27 around HJD~2458967.2 was due to weather conditions, manifest also in the reference lightcurves. The major event on 2020 April 26 detected also with TAOS was analyzed separately and not shown here.  }
 \label{fig:alldata}
\end{figure*}

\begin{figure}[htbp]
\centering
  \includegraphics[width=0.8\columnwidth,angle=0]{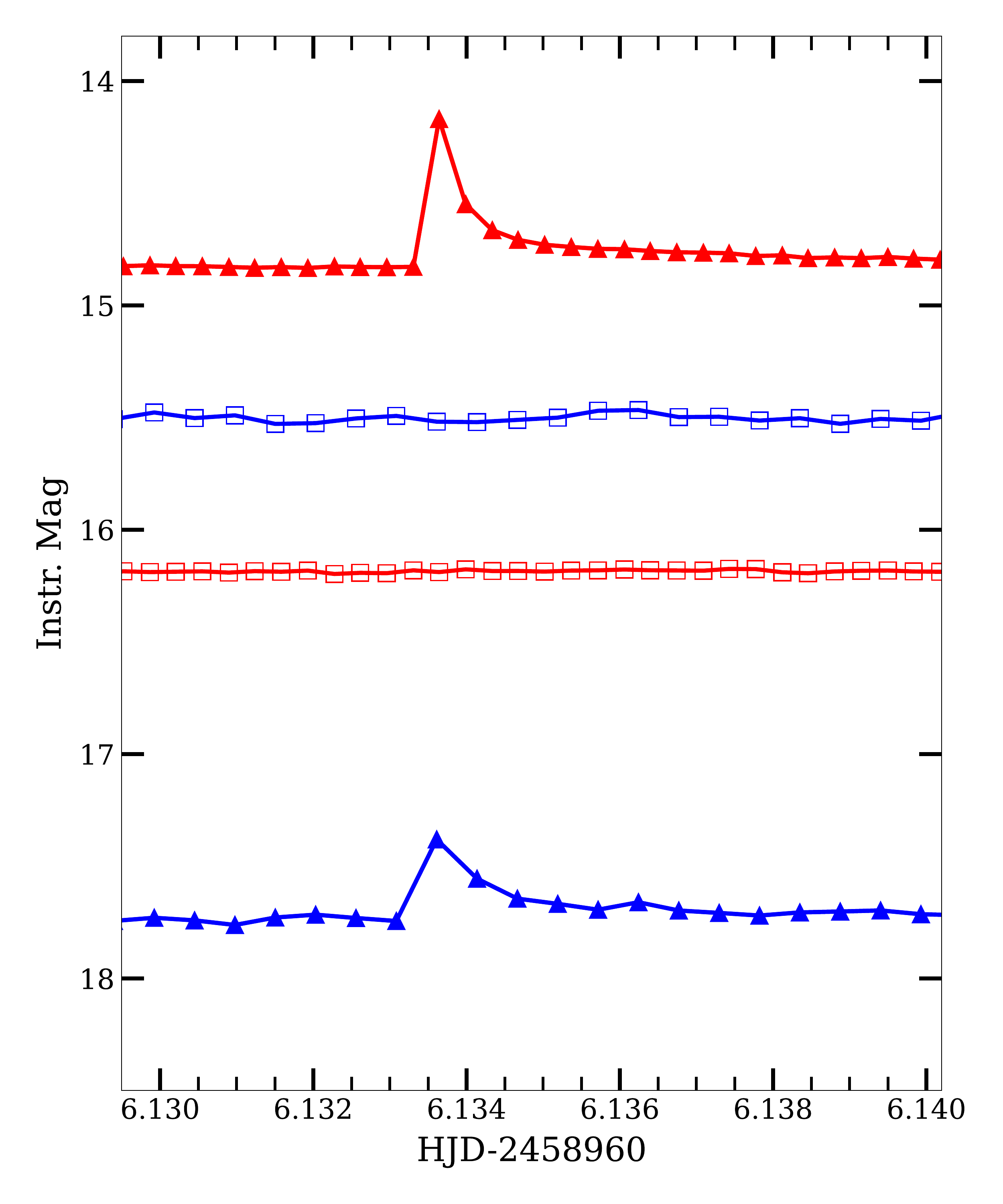}
   \caption{The major event detected on 2020 April 26 by both NOWT (in red) and TAOS (in blue).  For each telescope, the measured instrumental magnitudes of Wolf\,359 and of a comparison star are displayed to validate the variability.  Typical photometric errors are about 0.01~mag so smaller than the sizes of the symbols.
    }
  \label{fig:2telref}
\end{figure}

No standard star was observed and the brightness is referenced to that of the stellar quiescent state. In every case, two reference stars near Wolf\,359 in the same image frame were also measured to assess any variations due to the sky.   Figure~\ref{fig:alldata} exhibits the NOWT lightcurves obtained during the campaign with the individual flare events marked.  The major event detected on 2020 April 26 by both NOWT and TAOS telescopes is presented separately in Figure~\ref{fig:2telref}.  While the reference stars remained steady in brightness, Wolf\,359 experienced a brightening of $\sim0.65$~mag detected by NOWT, and $\sim 0.38$~mag detected by TAOS.  

Table~\ref{tab:lc} summaries the photometric measurements used to plot the lightcurves in Figure~\ref{fig:alldata}.  Columns 1, 2, and 3 list, respectively the calendar date, telescope, and the Heliocentric Julian Date (HJD) of the observation (middle of an exposure).  The remaining columns are magnitude and associated error of Wolf\,359, and of the two reference stars.

\startlongtable
\begin{deluxetable*}{lllllllll}
\tablecaption{Photometric data for Wolf\,359 and reference stars.
               \label{tab:photometry}}
\tablehead{
\colhead{Date} & \colhead{Telescope} & \colhead{Epoch} & 
\colhead{$m_\mathrm{W}$\tablenotemark{a}} & \colhead{$\sigma_{m_\mathrm{W}}$}\tablenotemark{a} & \colhead{$m_\mathrm{1}$\tablenotemark{b}} & \colhead{$\sigma_{m_\mathrm{1}}$\tablenotemark{b}} & \colhead{m$_\mathrm{2}$\tablenotemark{c}} & \colhead{$\sigma_{m_\mathrm{2}}$\tablenotemark{c}} \\
\colhead{} & \colhead{} & \colhead{HJD} & \colhead{mag} & \colhead{mag} & \colhead{mag} & \colhead{mag} & \colhead{mag} & \colhead{mag}
}
\startdata
2020 April 23 & NOWT & 2458963.1156348 & 14.7537 & 0.0016 & 16.1537 & 0.0032 & 16.1891 & 0.0033 \\
 & & 2458963.1169079 & 14.7529 & 0.0019 & 16.1874 & 0.0040 & 16.1823 & 0.0041 \\
 & & 2458963.1181114 & 14.7546 & 0.0019 & 16.1804 & 0.0040 & 16.1860 & 0.0040 \\
 & & 2458963.1184702 & 14.7455 & 0.0019 & 16.1843 & 0.0040 & 16.1740 & 0.0041 \\
 & & 2458963.1188406 & 14.7506 & 0.0019 & 16.1736 & 0.0039 & 16.1807 & 0.0040 \\
 & & 2458963.1192110 & 14.7538 & 0.0019 & 16.1769 & 0.0040 & 16.1817 & 0.0040 \\
 & & 2458963.1195697 & 14.7526 & 0.0019 & 16.1965 & 0.0040 & 16.1828 & 0.0040 \\
 & & 2458963.1199400 & 14.7478 & 0.0019 & 16.1916 & 0.0040 & 16.1856 & 0.0041 \\
 & & 2458963.1203104 & 14.7497 & 0.0019 & 16.1996 & 0.0041 & 16.1905 & 0.0041 \\
\enddata
\tablenotetext{a}{The variable $m_\mathrm{W}$ is the magnitude of Wolf 359.}
\tablenotetext{b}{The variable $m_\mathrm{1}$ is the magnitude of reference star 1.}
\tablenotetext{c}{The variable $m_\mathrm{2}$ is the magnitude of reference star 2.}
\tablecomments{Table 1 is published in its entirety in the machine-readable format.
      A portion is shown here for guidance regarding its form and content.}
      \label{tab:lc}
\end{deluxetable*}

In a total of 27 data hours, 13 flare events were identified visually in the lightcurves, in accord with the (non)variation of the reference stars at the same time.  The flare parameters for each event such as the peak amplitude, the rising and decay time scales were derived, from which the integrated total energy was computed.

\subsection{Flare Properties} 

A flare profile is parameterized with (1)~the epoch and amplitude of the observed peak in the lightcurve, and relative to the peak, (2)~the rising time scale, and (3)~the decline time scale. Usually (2), signifying the energizing process is relatively fast, whereas (3), relevant to the cooling mechanism drops off slower, typified by an exponential or a power-law decay. 

First, the observed lightcurve in unit of flux (count) $f(t)$ is subtracted of then divided by  the detrended quiescent stellar count (a linear fit to the flux away from an event) within the spectral filter $f_0$. To compute the released energy from such a normalized lightcurve, $ \Delta f/f_0 = (f(t)-f_0)/f_0)$, we follow the equivalent duration method described by \citet{ger72}.  The flare is approximated by a typical black-body of effective temperature $T_{\rm flare}$ of 9000~K \citep{moc80}, thereby having a bolometric luminosity of 
\[ 
  L_{\rm flare} = \sigma A_{\rm flare}T^4_{\rm flare}, 
\] 
where $ \sigma$ is the Stefan-Boltzmann constant, and $A_{\rm flare}$ is the flare area, related to the observed flare luminosity as  
\[
  L^\prime_{\rm flare} = A_{\rm flare }\int B_{\lambda} (T_{\rm flare})  \, R_\lambda d\lambda.
\]
Here $B_{\lambda} (T)$ is the Planck function and $R_\lambda$ is the spectral response function for which only the filter response is considered.  The observed photospheric luminosity of a star of radius $R_*$ is then 
\[
  L^\prime_* = \pi R_*^2 \int B_{\lambda} (T_*) \, R_\lambda d\lambda, 
\]

The ratio of $L^\prime_{\rm flare}$ to $L'_* $ is the amplitude, 
\[ 
  C^\prime_{\rm flare} = \frac{\Delta f}{f_0} = \frac{L^\prime_{\rm flare}}{L^\prime_{*}}, 
\]
and the area of the flare becomes
\[
  A_{\rm flare} = \pi R_*^2 \, C^\prime_{\rm flare} 
  \frac{\int B_{\lambda} (T_*) R_\lambda d\lambda}{\int B_{\lambda} (T_{\rm flare}) R_\lambda d\lambda},
\]
with the flare luminosity being computed as, 
\[
  L_{\rm flare} = \frac{1}{4} L_* \, \left(\frac{T_{\rm flare}}{T_*} \right)^4 C^\prime_{\rm flare}\, \frac{\int B_{\lambda} (T_*) R_\lambda d\lambda}{\int B_{\lambda} (T_{\rm flare}) R_\lambda d\lambda}
\]

Finally the total flare energy is estimated from the bolometric luminosity of the star multiplied by the equivalent duration.

\begin{equation} \label{eq:Energy computation}
  E_{\rm flare} = \alpha L_* \int \frac{\Delta f}{f_0} dt, 
\end{equation}
where $\alpha$ is the constant accounting for the correction for the black body assumption and the filter response. Taking the stellar temperature as 2900K~\citep{fuh05}, $\alpha$ of 0.11 for the standard Johnson-Cousins $R$ filter, and 0.05 for the $V$ filter, Equation~\ref{eq:Energy computation}, adopting $\log L_*/L_\sun = -2.95 \pm 0.05$ \citep{pav06}, leads to the derivation of the flare energy, $E_{\rm flare}$ released in an event.  Note that this method assumes the flare temperature to be constant throughout the event, and the behaviour of the flare is the same in all spectral bands.

Table~\ref{tab:events} summarizes the parameters of the 13 events including the superflare detected simultaneously by two telescopes.  For all events reported here, the rising time is less than $\sim30$~s, i.e., shorter than the cadence of each of the telescopes, so was not derived.   This rising/heating time scale is contrasted to those of several minutes among solar-type flares \citep{yan21}.  In our analysis, the lightcurve then takes a straight line from the stellar quiescent state, i.e., one data point prior to the peak. The date/time refers to the middle of the exposure within which a flare occurred.  The duration of an event is estimated from one data point prior to the peak to where the lightcurve falls below the uncertainty in $\Delta f/f_0$, typically 0.002 for NOWT and 0.01 for TAOS.  The determination of duration time therefore is somewhat subjective, but serves to gauge the relative time length of an event.  The events had energies ranging from  $\sim 3\times 10^{29}$~erg to $\sim 3\times 10^{31}$~erg, lasting for a couple of minutes to over 20 minutes.  Our campaign was not long enough to catch more powerful, presumably rarer events.  

We note that the term ``superflare'' is applicable to solar events, but not well defined for stars, whether it refers to total released energy (in unit of energy) or relative to stellar photospheric luminosity (in unit of power).  A superflare of solar-type stars releases $10^{33}$--$10^{38}$~erg \citep{sch00}, which given a typical duration of $\sim30$~min \citep{yan21} amounts to a ratio to stellar luminosity $10^{-4}~\mathrm{L}_\sun \lesssim L_\mathrm{flare} \lesssim 10^1~\mathrm{L}_\sun$. An M dwarf flare, on the other hand, gives out a total energy, $10^{31}$--$10^{34}$~erg, with the fast rotators liberating more, up to $10^{35}$~erg \citep{lin21}.  For the events reported here, the most energetic one has  $E\sim3\times10^{31}$~erg within $\sim25$~min, hence with $L_*=1.1 \times 10^{-3}\, L_\sun$ for the star, leading to $L_{\rm flare}/L_* \approx 0.5\%$.  The two-telescope event is powerful, having the rising and exponential decay time scales both less than about 30~s, with a peak amplitude comparable to the stellar flux, we refer this major event as a superflare.

\begin{deluxetable*}{ccCcc}
\tablecaption{Flare Event Parameters \label{tab:events}}
\tablewidth{0pt}
\tablehead{
\colhead{Date/Time} & \colhead{Amplitude} & \colhead{Energy} & \colhead{Time Constant} & \colhead{Duration} \\
\colhead{HJD$-2458960$} & \colhead{($\Delta f/ f_0 $)} & \colhead{(erg)} & \colhead{(s)} & \colhead{(s)} 
          }
\startdata
3.1974805 & 0.09 & 5.68  \times 10^{30} & 61  & 698   \\
4.1942563 & 0.05  & 1.48 \times 10^{30}  & 61  & 534 \\
4.2042208 & 0.02  & 4.46 \times 10^{30} & 457 & 1248  \\
5.1751989 & 0.03  & 3.47  \times 10^{30} & 15  & 712  \\
5.1937739 & 0.09  & 4.72 \times 10^{30} & 12 & 772 \\
6.1336424 & 0.82  & 3.28  \times 10^{31} & 23  & 1515 \\
6.1336110$^{a}$  & 0.41 &  3.26 \times 10^{31} & 43 & 1319\\
7.1918455 & 0.08  & 6.64 \times 10^{30} & 34  &745 \\
8.1159728 & 0.04  & 1.08 \times 10^{30} & 83  & 350  \\
8.1413180 & 0.03  & 2.87 \times 10^{29} & 35 & 126 \\
8.1512477 & 0.05  & 8.30\times 10^{29} & 98 & 254  \\
9.1471391 & 0.04  & 2.79\times 10^{29} & 23 & 126 \\ 
9.1739656 & 0.17  & 3.18 \times 10^{30} & 35  & 603  \\
9.1912327 & 0.18  & 3.38 \times 10^{30} & 47 & 539 \\
\enddata
\tablecomments{$^a$: Measured also by TAOS; all others by NOWT only
}
\end{deluxetable*}

\section{The Flare Rate of Wolf 359}

\begin{figure}[htbp]
\centering
    \includegraphics[width=0.8\columnwidth]{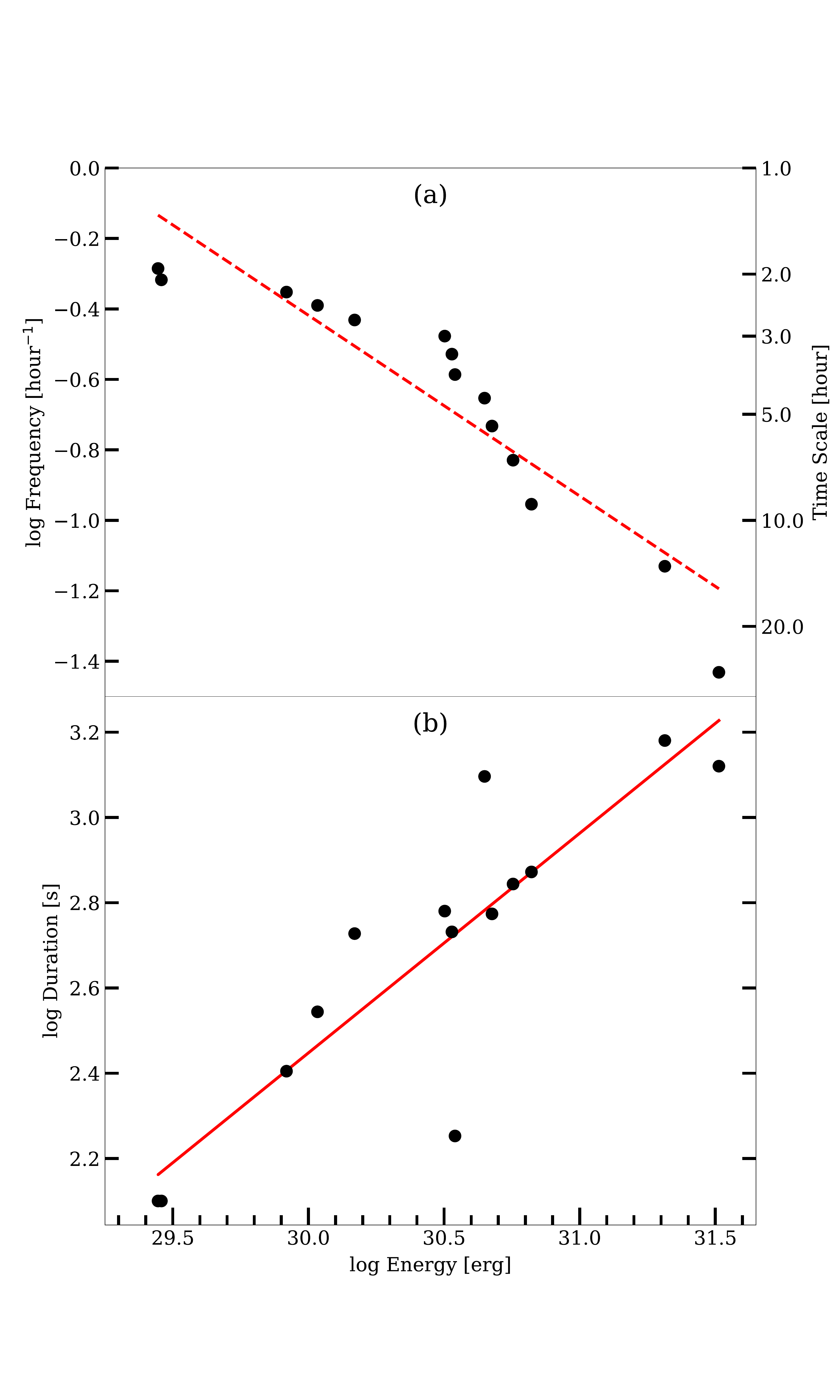}
  \caption{(a)~Cumulative frequency (or occurrence time scale) distribution and (b) duration versus flare energy of the Wolf\,359 flares reported in this work.  The dashed line in (a) is not meant for a fit but for illustration only because the data seem more complex than a linear fit.
    }
  \label{fig:ffd}
\end{figure}

The flare rate we detected, 13 flares in 27 hours, equivalent to about 48 events per 100 hours, is consistent with literature results \citep[e.g.,][47/100 in radio wavelengths]{nel79}.  In terms of energetics, Figure~\ref{fig:ffd} presents the cumulative frequency distribution of the flares of Wolf\,359 listed in Table~\ref{tab:events}.  Fitted with a power-law, $\log \nu \propto \beta \log E$, where $\nu$ is the flare frequency distribution with energy greater than $E$ \citep{hun12}, the index $\beta=-0.87\pm0.19$ if the data at the lowest-energy end, incomplete in our experiment, are excluded. Even with our limited number of events, spanning two orders of magnitude in energy, the distribution seems more complex than a single power-law, a conclusion that has been drawn for this star and for active M dwarfs in general \citep{lin21}.  By and large, Wolf\,359 produces a flare as powerful as $10^{30}$~erg approximately once every three hours.  While a fast rotating M dwarf like Wolf\,359 tends to produce frequent and powerful flares \citep{lin19}, it is not known whether the boosted magnetic activity of our target is linked to its own rapid spin alone or to the close-in planet with an enlarged emission volume.

A more energetic flare conceivably would take a longer time to dissipate.   Because a flare profile may be complex (e.g., more than a simple exponential or power-law decay, or multiple flares in the lightcurve segment), we correlate the duration of a flare with energy, exhibited also in Figure~\ref{fig:ffd}.  A log-log linear fit gives $t_{\rm flare} \propto E_{\rm flare}^{0.49\pm 0.08}$.  \citet{mae15} related the magnetic reconnection time scale to the Alv\'{e}n time and derived analytically for solar-type stars, $t_{\rm flare} \propto E_{\rm flare}^{1/3}$, to be comparable to their observed slope of $0.39 \pm 0.03$.  Such a duration-energy relation is also observed in the active dM\,4e dwarf GJ\,1243 \citep[][of a slope about 0.4, reading from their Figure~10]{haw14}, and in other M dwarfs \citep{rae20}.  Our slope is marginally steeper but, given the incompleteness for weaker events, should be overestimated.  If so, this indicates a similar energizing mechanism in M dwarfs as in solar-type stars.

\section{Intrinsic Flare Profile Derived from Two-Telescope Lightcurves} 

The simultaneous detection of the same event with different sampling functions allows us to break the parameter degeneracy and thereby to  infer the underlined flare profile.  By varying profile parameters, the lightcurve that would have resulted for a given sampling function of a telescope is compared with what was actually observed.  The profile that yields the least deviation in the chi-squared sense in the two lightcurves, is considered the ``best'' solution, closer to the ``truth'' than any individual observed lightcurve alone.  

\begin{figure}[htbp]
\centering
    \includegraphics[width=\columnwidth,angle=0]{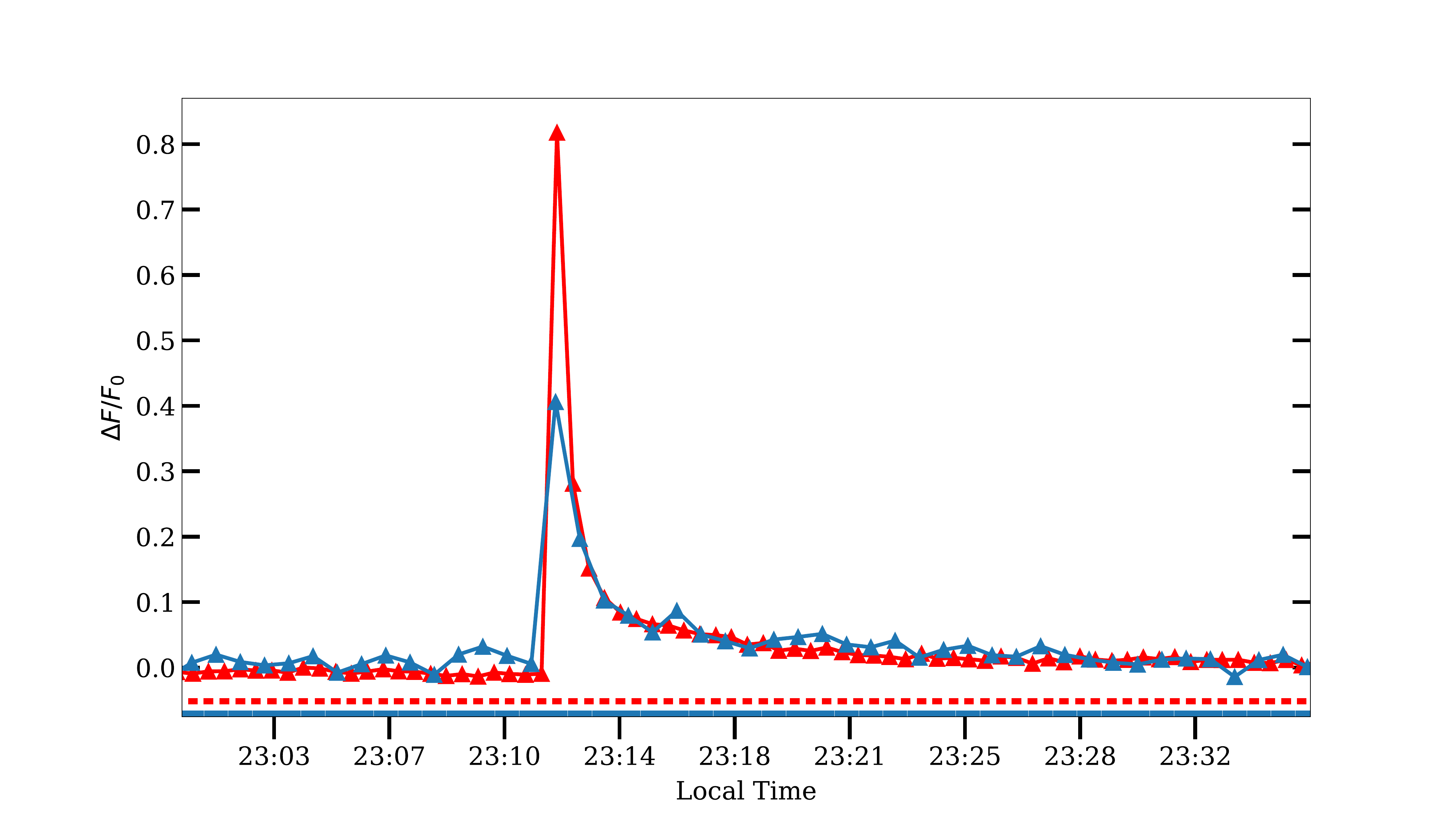}
  \caption{As in Figure~\ref{fig:2telref} for the superflare observed in 2020 April 26, but now each lightcurve has been detrended and rescaled relative to the quiescent stellar brightness, in flux/count unit instead of magnitude.  The horizontal segments mark the sampling function for NOWT (in red) and for TAOS (in blue), with the filled parts for exposure and open (or a dark bar for TAOS) for dead time intervals.
    }
  \label{fig:2lcs}
\end{figure}

The normalized lightcurves by the two telescopes are displayed in Figure~\ref{fig:2lcs}.  Note that NOWT detects a relative amplitude of $\sim0.8$, whereas TAOS detects an amplitude of $\sim0.4$.  There is a caveat that, because of the different filters used, $f_0$ differs for each telescope.  Conceivably, the star itself is red in color, but the flare should be relatively blue, but here we assume the same $f_0$ to normalize each lightcurve.  The peak of the flare was sampled differently: TAOS detected it at HJD of  2458966.133611, whereas NOWT detected it at HJD of 2458966.1336424, with an offset of about 2.7~s.

The sampling effects on an intrinsically continuous function include the integration time (averaging the signal), dead time (no signal), and the lapse, i.e., the offset time between the peak relative to the sampling window.  It is this offset and finite sampling time that smear off the peak of the flare and distort the shape of a lightcurve, accounting for different flare statistics between the long-cadence versus short-cadence K2 lightcurves \citep[e.g.,][]{rae20,lin21}. A grid of event parameters were used to compute the simulated lightcurves with steps of 0.01 in the peak amplitude and 1~s in decay time scale, chosen as the appropriate step parameters with extensive simulations.  For the decay portion, different models were exercised:~a single exponential function ($A\, e^{-t/\tau_1}$), a double-exponential function ($A\, e^{-t/\tau_1}+B\, e^{-t/\tau_2}$), and a power-law fall-off ($A\, t^{-\gamma}$), where $A$ and $B$ are amplitudes, $t$ is time, $\tau_1$ and $\tau_2$ are correspondent exponential time constants, and $\gamma$ is the power-law exponent index.  In each case, the computed lightcurve according to a specific sampling function was  compared with the actual observed one (``observed'' minus ``computed'', ``$O-C$'')  to evaluate the chi-squared value  (sum of $(O-C)/O$).

\begin{figure}[htbp]
\centering
 \includegraphics[width=\columnwidth]{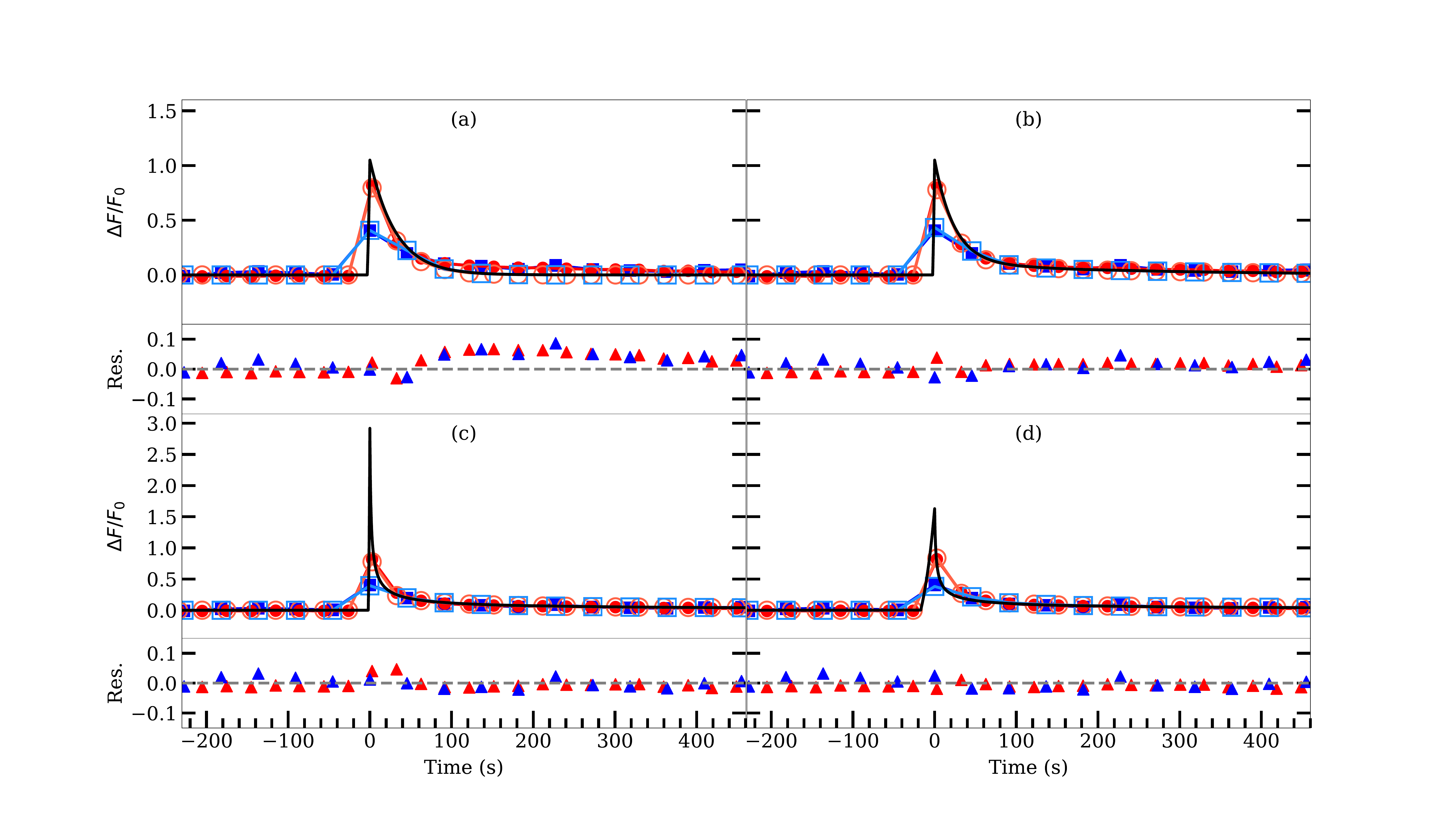}
  \caption{The observed (filled symbols, NOWT in red and TAOS in blue) versus computed (open symbols) lightcurves for (a)~a single-exponential, (b)~a double-exponential, (c)~a power-law decay function with an instantaneous impulse rise, and (d)~a power-law decay function with a finite-time rise (16~s).  For each model, the analytic function is represented as a black curve, and the residuals ($O-C$, the observed minus the computed) are also shown.  For (c) the peak amplitude is $\Delta F/F_0 = 2.92$, while in (d) it is $\Delta F/F_0 = 1.63$.  
    }
  \label{fig:allin1}
\end{figure}

Figure~\ref{fig:allin1} presents the best-fit results, whose parameters are summarized in Table~\ref{tab:model}.  The double-exponential model, adding one more degree of freedom in the fitting, gives an over-all better account than the single-exponential function of the fading part of the lightcurves, judged by the residual $\chi^2$.  This is consistent with the time-resolved flares of another eruptive red dwarf, GJ\,1243 \citep{dav14}, and supports the notion of possibly more than one cooling mechanism \citep[radiative and conductive,][]{ben10}. 

For the power-law model, we present two cases, one with an instantaneous impulse rise, and the other with a finite-time rise.  The latter is more realistic, but our data could not constrain the rising time scale.  Therefore an exponential rise of 16~s is adopted here as an example, for which a peak of 1.63 times of the stellar brightness is required to fit the data, whereas for an instantaneous rise, the peak would have been 2.92 then falling off faster (with a slightly larger value of $\gamma$).  In general a power-law decay requires a higher peak than an exponential model, leading to an elevated total flare energy.  In our data, the impulse plus the rapid decay portion of the lightcurve spans no more than a few data points.  This means that a higher time resolution is needed than reported here in order to distinguish one cooling function from another.  

\begin{deluxetable*}{ccccc cCC}
\tablecaption{Best-fit Model Parameters \label{tab:model}}
\tablewidth{0pt}
\tablehead{
\colhead{Model} & \colhead{A} & \colhead{B} & \colhead{$\tau_1$}  & \colhead{$\tau_2$} & \colhead{$\gamma$} &  \colhead{Energy (erg)} & \colhead{$\chi^2$} 
           }
\startdata
One-exponential & 1.05 &      & 32 &     &      & 5.97 \times 10^{30} & 0.014\\
Two-exponential & 0.94 & 0.11 & 24 & 234 &      & 8.41 \times 10^{30} & 0.007 \\
Power law       & 2.92 &      &    &     &  0.7 & 1.88 \times 10^{31}                 & 0.011 \\
Power law       & 1.63 &      &    &     &  0.6 & 2.06 \times 10^{31} & 0.004 \\
\enddata
\end{deluxetable*}

\section{Implication for Flare Observations }

\begin{figure}
    \includegraphics[width=\columnwidth,angle=0]{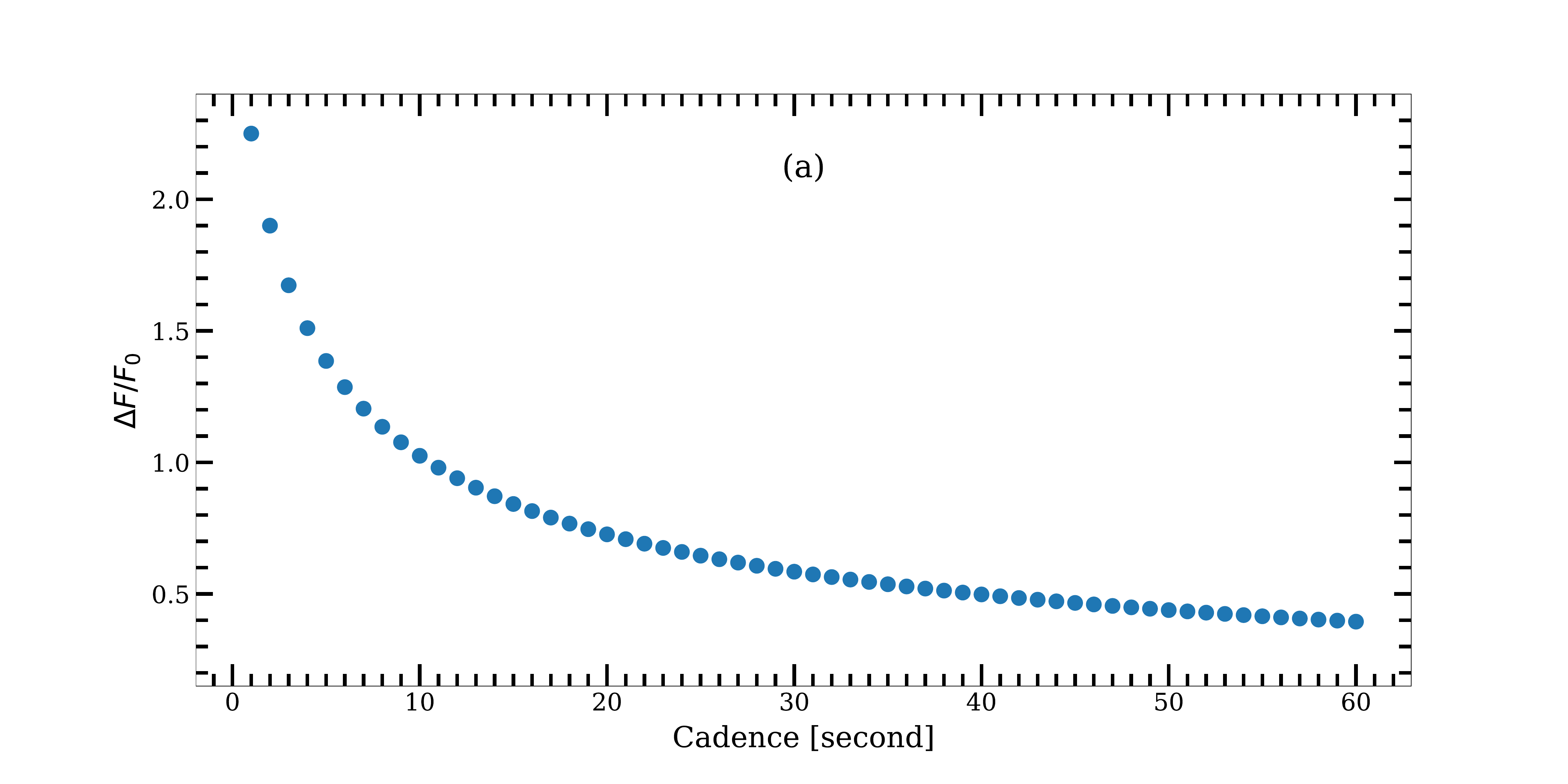}
    \includegraphics[width=\columnwidth,angle=0]{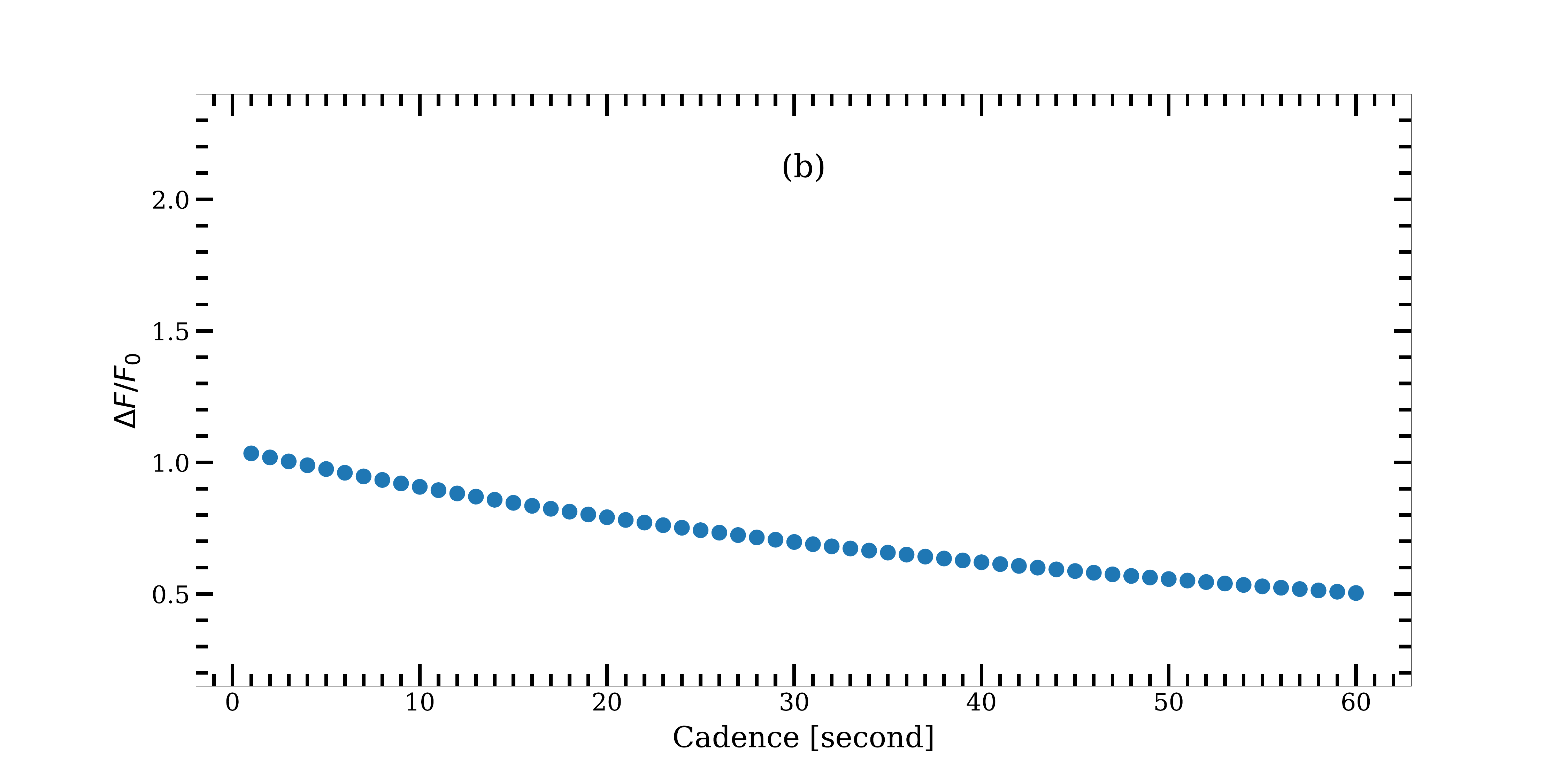}
    \caption{Effect of observing cadence of a continuous (a)~power-law, or (b)~a double-exponential function.    }
\label{fig:cadenceEffect}
\end{figure}

A flare event, detected either visually or by an algorithm (e.g., to recognize in a lightcurve an abrupt rise followed by a few data points above quiescence) is characterized by the amplitude and duration, from which the total energy is derive.  The fact that the true superflare event reaches to at least 1.6 times of stellar brightness, while the observed lightcurves peak, respectively, at 0.8 and 0.4 times, as demonstrated in this work, manifests how the sampling function affects the amplitude of an observed flare. The experimental design to detect sporadic stellar flares hence pertains to an integration time as short as possible so as to resolve the flare profile given the kind of flare events targeted for detection, while commensurate with sufficient signal-to-noise. Figure~\ref{fig:cadenceEffect} plots how cadence affects the detected amplitude of the superflare event reported here, for which the peak of 1.6 times of stellar brightness would be degraded quickly; e.g., with a 30-s integration the detected peak drops to less than 70\%.  This applies only to the specific event detected on 2020 April 26 for Wolf\,359, but serves to demonstrate vividly the essence of fast sampling.  The lesson is while the total energy of a flare can be reasonably estimated with a single data set, different samplings of a flare is necessary to derive the true profile in order to distinguish the heating and cooling processes.   One improvement of our experiment, other than with larger telescopes to afford faster cadences, is to measure the same event at different passbands, or better yet with spectroscopy, thereby diagnosing the temperature variation during the flare. 

Stellar flare activity may be elevated if the field lines have an external source to anchor to, be it a circumstellar disk, a companion, or an exoplanet, increasing the magnetic filling factors hence the emitting volume than by surface starspot pairs or coronal loops \citep{ben10}.  In stars like Wolf\,359 there may well be a combination of solar-type surface flares plus inflated star-planet events.  Long-term high cadenced monitoring observations are called for to derive any possible rotation or orbital periodicity.

\begin{figure}
  \includegraphics[width=\columnwidth,angle=0]{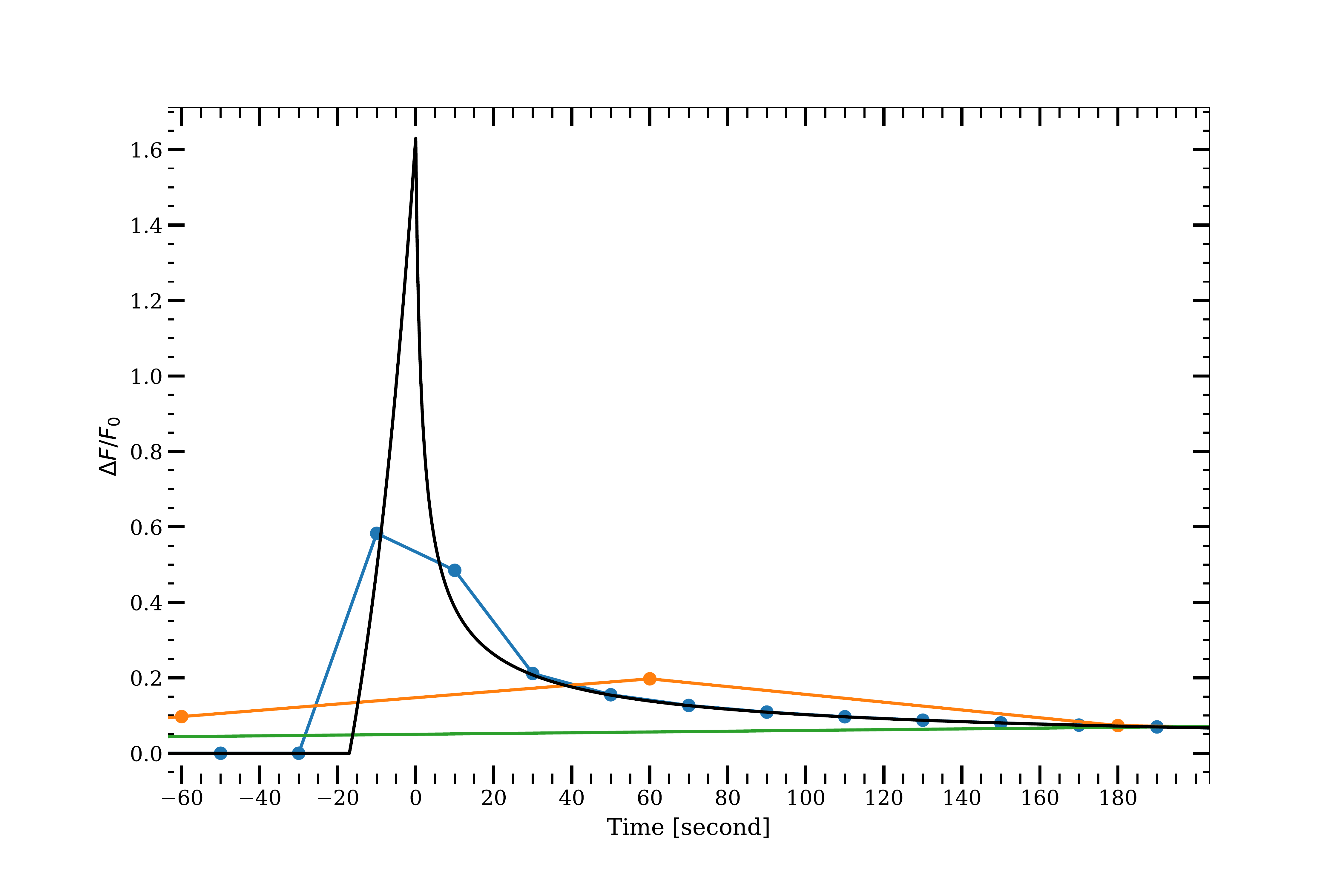}
  \includegraphics[width=\columnwidth,angle=0]{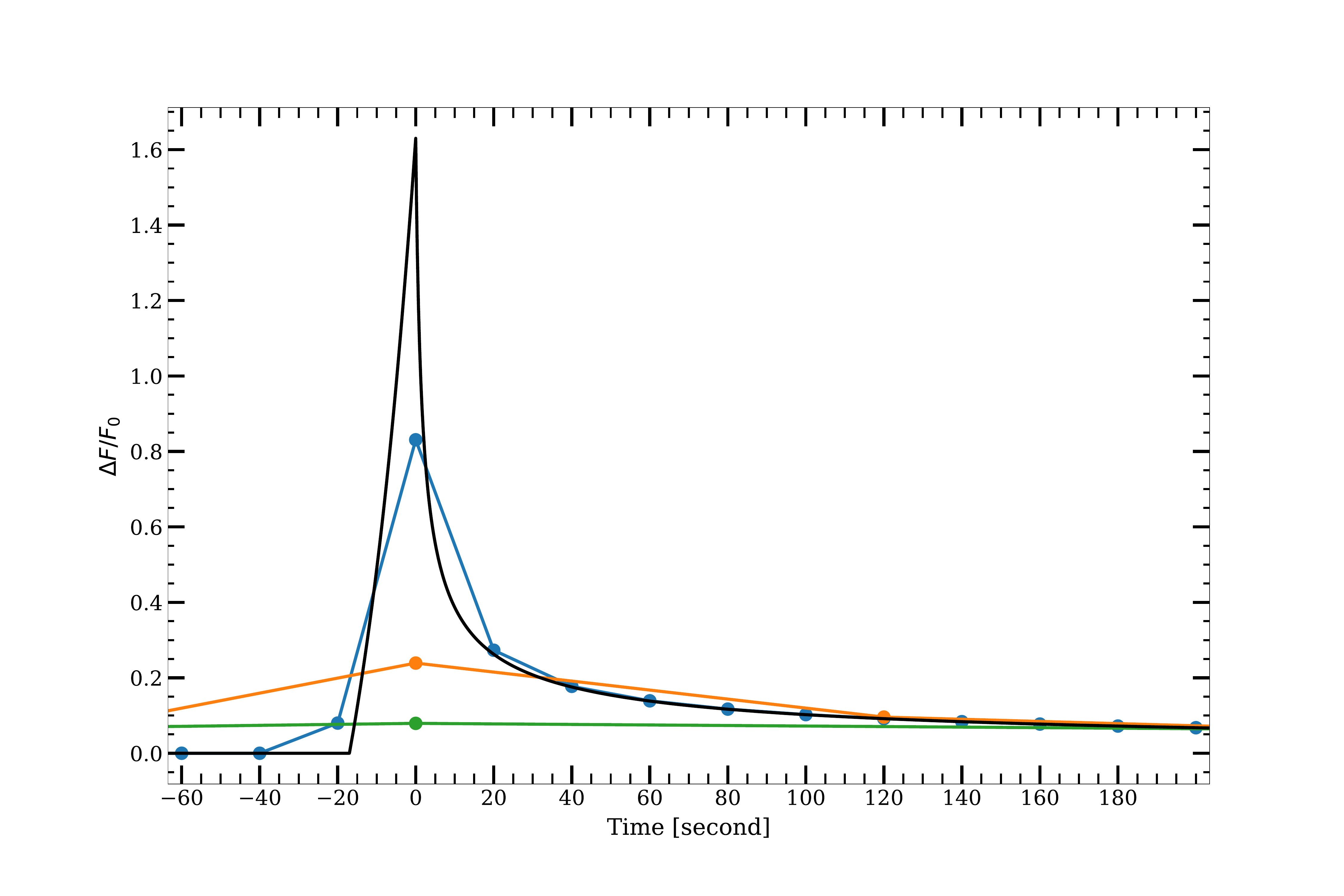}
\caption{The computed lightcurves of the intrinsic flare profile with a peak of 1.6 times of the stellar brightness (Figure~\ref{fig:allin1}(d)) sampled at a cadence of 20~s (in blue), 2~min (in orange), and 20~min (in green) (a) with a zero phase lag, or (b) with a 0.5 phase lag.  }
\label{fig:tess}
\end{figure}

As in the case of Kepler/K2, the Transiting Exoplanet Survey Satellite \citep[TESS,][]{ric15} provides data as useful for stellar flare research as in the primary science in exoplanets, particularly if complemented with ground-based observations of high sampling rates, \citep[e.g.,][]{how20}.   TESS are monitoring some bright stars of three different cadences: 20~s, 2~min, and 10~min.  Figure~\ref{fig:tess} illustrates how these three sampling rates would have detected the April 26 event.  Here a peak amplitude of 1.6 times of the stellar brightness is adopted (c.f., Figure~\ref{fig:allin1}(d)) with a zero phase lag, i.e., with the peak coinciding with the start of the sampling window, versus with a 0.5 phase lag.  One sees that for this particular flare only the shortest (20~s) cadence, similar to the data reported here, can resolve the profile, with a phase-dependent amplitude of 0.6 or 0.8, respectively, but neither the 2~min (for selected targets) nor the 10~min (for the whole frame) cadence can.

\begin{figure}
    \includegraphics[width=0.75\columnwidth,angle=0]{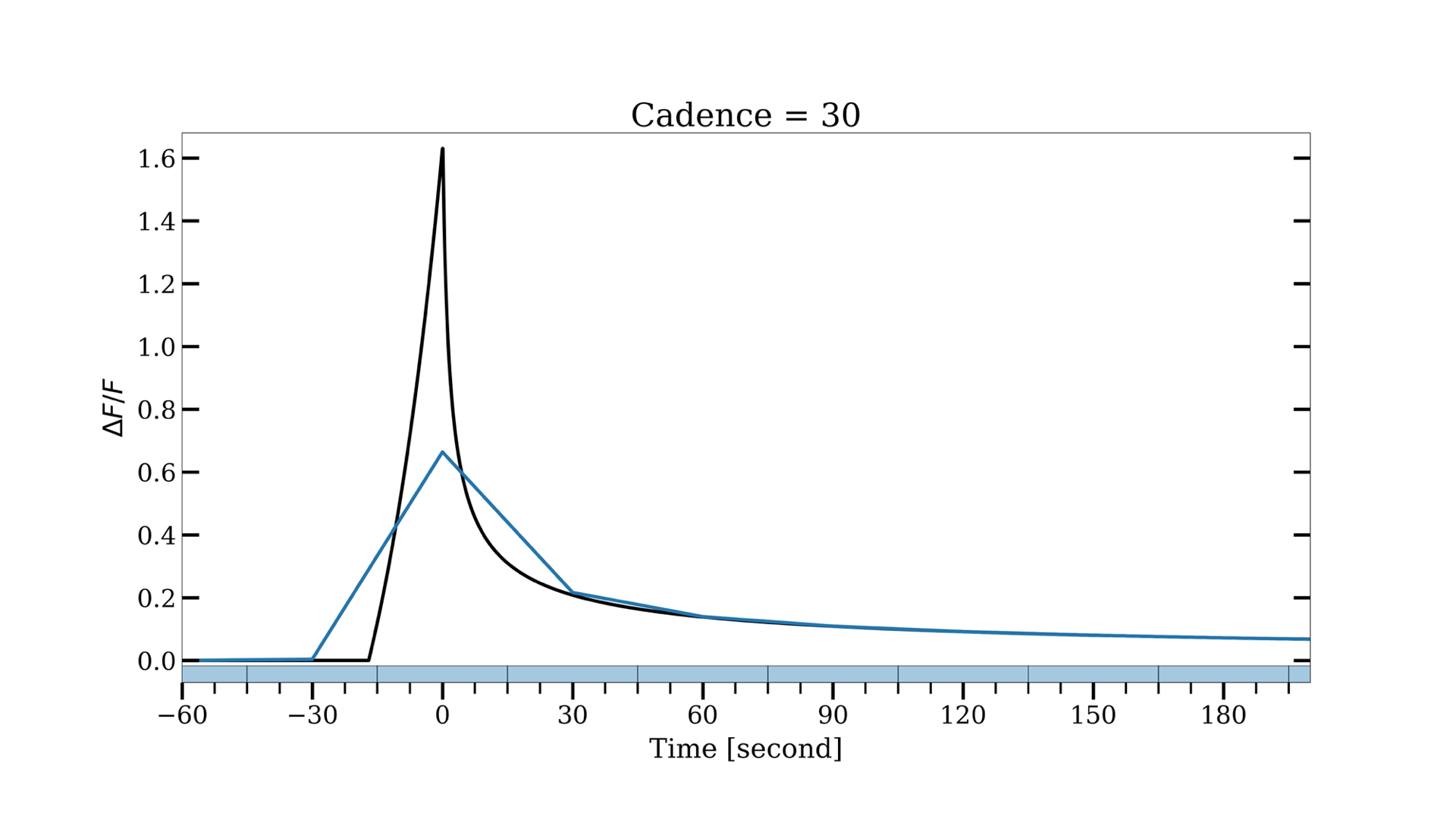}
\caption{An illustration of the effect of discrete sampling of a continuous flare profile.}
\end{figure}

In summary, our photometric monitoring of the red dwarf Wolf\,359 in 2020 April detected, in 27 data hours, 13 flare events with released energy in the range $\approx 3\times 10^{29}$--$3\times 10^{31}$~erg, consistent with the flare occurrence rate for this star reported previously in the literature.  For any single-telescope data, the peak and energy are underestimated as the result of sampling by finite integration time with a phase lapse.  A major flare was detected simultaneously by two telescopes on 2020 April 26, for which the underlying flare profile is estimated.  The profile parameters are model dependent, but the ``true'' flare amplitude might reach as high as 1.6 times of the quiescent stellar flux, whereas the two telescopes detected a peak level of 0.8 and 0.4, respectively, with the total released energy nearly four times as large.

\begin{acknowledgments}

We thank the referee for constructive comments to improve the quality of the paper.  We are grateful to Xinjiang Astronomical Observatory for support in installation and operations of the TAOS telescopes in Qitai.  The XAO authors acknowledge the financial aid from National Natural Science Foundation of China under grant U2031204.  The relocation of the two TAOS telescopes was funded by the grant MOST\,105-2119-M-008-028-MY3.  The work at NCU is financially supported in part by the grant MOST\,109-2112-M-008-015-MY3.  

\end{acknowledgments}

\bibliographystyle{aasjournal}
\bibliography{main}{}

\end{document}